\documentclass[12pt,a4paper]{article}
\usepackage[utf8]{inputenc}
\usepackage[T1]{fontenc}
\IfFileExists{lmodern.sty}{\usepackage{lmodern}}{}
\usepackage[british]{babel}
\usepackage{amsmath,amssymb,amsthm,mathtools}
\usepackage{graphicx}
\usepackage{booktabs,tabularx,multirow,longtable,array}
\usepackage{caption}
\usepackage{float,placeins}
\usepackage{enumitem}
\usepackage{setspace}
\usepackage[round,authoryear]{natbib}
\usepackage[margin=2.5cm]{geometry}
\usepackage[hidelinks]{hyperref}
\newcommand{\E}{\mathbb{E}}
\theoremstyle{definition}

\title{\textbf{A Three-Factor Hull--White Scenario Generator for Macro-Financial Stress Forecasting}}
\author{Ahmad Qadi\thanks{Corresponding author. Email: \texttt{ahmad.qadi.22@ucl.ac.uk}.}\quad Thomas Schr\"{o}der\quad Francesca Medda\\[0.6em]
\normalsize UCL Institute of Finance and Technology\\ \normalsize University College London, London, United Kingdom}
\date{}

\begin{document}
\maketitle
\thispagestyle{empty}

\begin{abstract}
\noindent This paper builds a scenario engine that maps a yield-curve path into a distribution of macro-financial outcomes. A three-factor Hull--White model generates arbitrage-free curves whose factors are simulated under the physical measure, with an estimated and reported market price of risk. A Minnesota-prior VARX maps the simulated curves into weekly equity returns, the federal funds rate, the Weekly Economic Index, inflation and a bank Marginal Expected Shortfall proxy, with a sign-restricted policy shock adapted to weekly data. Evaluated out of sample from 2017 to 2024, the term-structure block has the lowest forecast error in 13 of 20 maturity and horizon cells, and the density forecasts are well calibrated for the continuously moving variables. At weekly frequency the curve carries no incremental information about these variables once their own lags are controlled for, so the engine's value lies in conditioning on a specified curve path, which no benchmark can do.
\end{abstract}



\newpage
\section{Introduction}
\label{c2:sec:introduction}

When a central bank raises rates to slow inflation, it also raises funding costs, narrows bank interest margins and weakens leveraged balance sheets \citep{AdrianShin2010, BorioZhu2012}. The channel through which both effects travel is the term structure of interest rates \citep{BernankeBlinder1988, GertlerKaradi2015}. A supervisor who asks what a given path for rates would mean for activity, inflation and bank fragility is therefore asking for a joint distribution of those outcomes, conditional on that path.

No standard tool produces it. Macroeconomic models trace the effects of policy shocks but impose no arbitrage-consistent structure on the curve. Term-structure models generate coherent curves but connect to nothing downstream. Regulatory stress tests use a handful of hand-built scenarios rather than a distribution over plausible ones. This paper builds the missing piece: a transparent, reproducible engine that converts a yield-curve path into a conditional macro-financial distribution, and evaluates it to the standard normally applied to forecasting models.

The engine has two stages. The first is a three-factor Gaussian Hull--White model \citep{HullWhite1990}. Its curves are priced under the risk-neutral measure, so they are arbitrage-free and reproduce the observed discount curve exactly, but the factors that drive them are simulated under the physical measure, estimated from fifteen years of weekly curve history. An explicitly estimated market price of risk links the two. This is the paper's central methodological step: it lets one model be both arbitrage-free and a credible description of how yields move, and it puts the term premium embedded in every scenario on the page as three numbers. The second stage is a Minnesota-prior VARX \citep{Doan1984, Litterman1986} in five weekly variables: equity returns, the change in the federal funds rate, the Weekly Economic Index of \citet{LewisEtAl2022}, year-on-year CPI inflation and a bank-sector Marginal Expected Shortfall proxy. The three yield-curve principal components enter as exogenous conditioning variables, which is what allows any curve path to be imposed and propagated. A contractionary policy shock is identified by sign restrictions \citep{Uhlig2005, RubioRamirez2010}, with a relevance filter and magnitude caps that address a problem specific to weekly monetary VARs.

The paper makes four contributions.

First, it turns a pricing model into a physical-measure scenario generator with an explicit and inspectable risk premium. Applied work in this area often leaves the choice of measure unstated; here it is estimated, reported and shown to be economically ordinary.

Second, it supplies two identification refinements that make sign-restricted monetary shocks usable at weekly frequency, where the policy rate is unchanged in most weeks. Without them, rotations that carry almost no policy-rate content dominate the identified set once it is normalised.

Third, it evaluates every component out of sample. Over 29 to 31 expanding-window origins from 2017 to 2024, the term-structure block attains the lowest RMSE in 13 of 20 maturity--horizon cells, with a tie in a fourteenth, and the transmission layer produces well-calibrated density forecasts for the three continuously moving variables. Institutional frameworks of this kind, such as the Bank of England's RAMSI and the ECB's BEAST, are not publicly evaluated against simple statistical baselines; this one is.

Fourth, the evaluation establishes an empirical result of independent interest. The predictive content of the yield curve for real activity is one of the most robust findings in macroeconomics at quarterly and annual horizons. At weekly frequency, and once each variable's own lags are controlled for, it is not there: three independent tests agree that the curve adds no measurable information, in the conditional mean or in the full predictive distribution. That result is useful to anyone building high-frequency monitoring around curve movements, and it clarifies what the framework is for. Its value lies in the conditioning channel, which a supervisor can use to impose a parallel rise, an inversion or a steepener, paced over a chosen number of weeks, and obtain a coherent distribution under it, not in accuracy gains flowing through that channel. No univariate or no-change benchmark can produce such a conditional distribution, because none has a yield-curve input.

The structural block adds a transmission result that is informative in its own right: MES and a bank--market covariance proxy respond in opposite directions to the same tightening shock, which shows that a tail-risk level measure and a co-movement measure are complements rather than substitutes.

Section~\ref{c2:sec:literature} places the paper in the literature. Section~\ref{c2:sec:methodology} sets out the data, the factor extraction, the G3++ calibration, the VARX and the identification. Section~\ref{c2:sec:results} reports the results, Section~\ref{c2:sec:policy} draws out the implications for practice, and Section~\ref{c2:sec:conclusion} concludes.

\section{Literature Review}
\label{c2:sec:literature}

Four literatures bear on this paper: term-structure modelling, the identification of monetary-policy shocks, the measurement of systemic risk, and probabilistic stress testing. Each is well developed on its own. What is missing is a framework that joins them.

The first is the modelling of the term structure. Stochastic interest-rate modelling begins with \citet{Vasicek1977} and \citet{Cox1985}. \citet{HullWhite1990} let the mean-reversion target vary with time, $dr_t = a(\theta(t) - r_t)\,dt + \sigma\,dW_t$, so that the model fits the observed yield curve exactly. Adding further Ornstein--Uhlenbeck factors gives the G2++ and G3++ models, which capture slope and curvature as well as level, matching the finding of \citet{Litterman1991} that three factors explain almost all of the variation in U.S. Treasury bond returns \citep[see also][]{DaiSingleton2000, DieboldLi2006}. Three factors suffice at the weekly frequency studied here. Recent work adds deterministic cycles: \citet{Jha2025SinusoidalHW} lets the mean-reversion speed of a single-factor Hull--White model vary sinusoidally to capture multi-decade periodicity in Treasury yields, which is not relevant at short and medium horizons.

How the model is calibrated depends on what it is for. Most Hull--White applications fit option prices under the risk-neutral measure, which is right for pricing derivatives but not for describing where yields are expected to go, since the drift of each factor differs between the two measures by the market price of risk \citep{Duffee2002}. Scenario generation needs the physical measure. This paper keeps both: curves are priced risk-neutrally, so they are arbitrage-free, and the factors are simulated under the physical measure, with an explicit market price of risk joining the two. Section~\ref{c2:subsec:hull_white} sets this out.

A related literature connects yields to the macroeconomy, and it runs in the opposite direction from this paper. \citet{AngPiazzesi2003} embed macro factors in a no-arbitrage VAR for yields; \citet{DieboldRudebuschAruoba2006} let level, slope and curvature interact with macroeconomic states in both directions; \citet{BauerRudebusch2020} decompose long-term yields into expectations and a term premium anchored to slow-moving trends. A separate strand, from \citet{EstrellaHardouvelis1991} to \citet{ChauvetPotter2005}, uses the yield spread to predict recessions. The present framework reverses the direction of conditioning: yields drive macro variables, because the term-structure model is the upstream scenario generator. The objective is also different. Rather than finding the single most informative yield variable, the aim is to generate a distribution of coherent curves and propagate it through a transparent transmission layer. Treating the curve as exogenous is a design choice for that purpose, not a claim that macro variables never feed back into yields.

The second literature concerns how monetary policy reaches the economy and the banking system, and how a policy shock can be identified. Policy operates through several channels. Communication moves the whole curve without a rate change \citep{Gurkaynak2005}. Tighter policy reduces the supply of intermediated credit, especially from weakly capitalised institutions \citep{KashyapStein2000, BernankeGertlerGilchrist1999}, as the repricing of duration risk in March 2023 illustrated. Prolonged low rates build vulnerabilities that surface when rates normalise \citep{BorioZhu2012, DellAriccia2017}, and the shape of the curve matters in itself because flattening compresses net interest margins \citep{English2018}. Estimating these effects requires an exogenous policy shock. Structural VARs moved from recursive identification \citep{Christiano1999} to sign restrictions \citep{Uhlig2005, RubioRamirez2010}. High-frequency identification isolates surprises in narrow windows around FOMC announcements \citep{Kuttner2001, Gurkaynak2005, JarocinskiKaradi2020} and uses them as external instruments \citep{MertensRavn2013, StockWatson2018}. That route is closed here: FOMC meetings occur roughly every six weeks, so weekly aggregates of the surprises are zero most of the time and first-stage relevance fails \citep{StockYogo2005}. Sign restrictions are therefore used, with two refinements developed in Section~\ref{c2:subsubsec:relevance_filter}.

The third literature supplies the measures of systemic risk that a transmission model should carry. Marginal Expected Shortfall is the expected equity loss of an institution when the market is in its lower tail, $\text{MES}_{i,t}^{(q)} = \E[-R_{i,t} \mid R_{m,t} \leq \text{VaR}_{m,t}^{(q)}]$ \citep{AcharyaEtAl2017}. CoVaR is the system's Value-at-Risk conditional on one institution being in distress \citep{AdrianBrunnermeier2016}. SRISK combines MES with size and leverage to estimate the capital shortfall a crisis would open up \citep{BrownleesEngle2017}. SRISK is unsuitable at weekly frequency because it depends on quarterly balance-sheet data, so its weekly differences are dominated by accounting jumps. MES updates continuously through equity returns and is the measure used here. A rolling bank--market covariance proxy, inspired by CoVaR but not the quantile-regression estimator of \citet{AdrianBrunnermeier2016}, is retained as a robustness check and turns out to be informative in its own right (Section~\ref{c2:subsec:results_irf}). Tightening is known to lower bank stock prices \citep{Ammer2010, GertlerKaradi2015}, more so for banks with greater interest-rate exposure \citep{English2018}.

The fourth literature is stress testing, which is where the first three would be put to use. Supervisory exercises such as CCAR and the ECB's stress tests evaluate capital under a few hand-built adverse scenarios, an approach argued to understate tail risk because it conditions on a few paths rather than a distribution \citep{Schuermann2014}; the Federal Reserve's CLASS model \citep{Hirtle2016} shares this design. Simulation addresses the problem by generating many trajectories, and the difficulty it faces is coherence: drawn naively, paths can pair sharp rate rises with implausible inflation or asset-price behaviour. The two-stage design here handles this structurally, because the term-structure block is arbitrage-free by construction and the macro-financial block is anchored by an estimated VARX.

Two institutional frameworks operate on this logic at scale. The Bank of England's RAMSI couples macro-financial dynamics to a bank balance-sheet block \citep{Alessandri2009}; the ECB's BEAST links a macro-financial engine to bank-level credit risk and solvency \citep{Budnik2020}. Both are far richer than the framework here in their treatment of individual institutions and neither is a competitor to it. They differ in two respects. In their published supervisory applications the rate-side input is a deterministic baseline-plus-adverse path rather than a distribution of curves, although BEAST does support stochastic simulation; and neither is publicly evaluated against simple statistical baselines. The contribution here is complementary: a scenario-generation and transmission layer small enough to be evaluated formally, which could sit upstream of such models. The BIS report on macro-financial stability frameworks \citep{BIS2022} lists among the open conceptual challenges the methods for identifying the build-up of vulnerabilities in real time and linking them to models for policy analysis, and the need to assess how a change in policy settings would alter the risk outlook. A transparent engine that maps a rate path into a distribution of macro-financial outcomes is one piece of that toolkit, and it is the piece this paper supplies.

No existing framework combines an arbitrage-consistent term-structure model simulated under the physical measure with an explicit market price of risk, a structural macro-financial VARX, and formal out-of-sample evaluation of both point accuracy and density calibration. That combination is what this paper provides.

\section{Methodology}
\label{c2:sec:methodology}

\subsection{Data}
\label{c2:subsec:data}

The analysis combines macro-financial series from January 2008 with a London Stock Exchange Group (LSEG) zero-coupon yield-curve panel from June 2010. The earlier macro start date supports lagged and rolling variable construction; the VARX sample is governed by zero-curve availability and runs from July 2010 to December 2024 after Friday end-of-week alignment and differencing.

\paragraph{Yield-curve data.} The term-structure block uses LSEG U.S. dollar zero-coupon curves, a LIBOR swap zero-coupon curve up to June 2020 and a SOFR OIS curve thereafter. The switch takes effect on 19 June 2020, three years before LIBOR cessation. Because the two curves differ in level by the LIBOR--OIS basis, the splice week records a downward jump across the whole curve of $-23.8$, $-23.2$, $-26.4$, $-28.8$ and $-30.5$\,bp at the 1Y, 2Y, 5Y, 10Y and 30Y pillars, three to four times the weekly zero-rate volatility reported in Table~\ref{c2:tab:g3pp_vol_fit}. The jump is an artefact of the change of curve, not a market move, and it is kept out of every estimate that uses weekly changes: the splice-week factor change is set to zero in the differenced series, which is equivalent to level-adjusting the SOFR curve to the LIBOR curve at the splice date for that purpose, and the transition across the splice is excluded from the factor-dynamics and volatility estimates. Levels are used as observed. Curves are retrieved at weekly Friday valuation dates, with holidays backtracked to the latest business date. Discount factors are converted into continuously compounded zero rates and interpolated onto fixed 1Y, 2Y, 5Y, 10Y and 30Y pillars.

\paragraph{Macro-financial variables.} The VARX endogenous block contains five weekly variables, all in first differences for stationarity: the S\&P~500 weekly log return (\texttt{mkt\_ret}); the weekly change in the effective federal funds rate, in decimals (\texttt{dffr}); the weekly change in the Weekly Economic Index of \citet{LewisEtAl2022} (\texttt{dwei}); the weekly change in year-on-year CPI inflation, forward-filled from monthly releases (\texttt{dinf}); and the weekly change in the market-cap-weighted bank-sector MES (\texttt{d\_mes\_proxy}).

Two of these series move only intermittently at weekly frequency. The federal funds rate is unchanged in most weeks between FOMC decisions, and CPI is forward-filled from monthly releases, so its weekly change is zero in roughly three weeks out of four. This is a consequence of matching the frequency of the yield-curve panel, and it has two implications that recur later. A no-change benchmark is close to optimal for these two variables by construction, so the forecast comparisons in Section~\ref{c2:subsec:results_phase8} are most informative for the three continuously moving series; and the small residual variance of the policy-rate equation is what makes the relevance filter of Section~\ref{c2:subsubsec:relevance_filter} necessary.

\paragraph{Systemic-risk construction.} MES is constructed from daily equity returns for 19 large U.S. banks.\footnote{Tickers excluded for short histories or data-quality reasons: ALLY, CFG, SYF, DFS, SIVB. SVB is excluded because its March 2023 failure produces discontinuous balance-sheet jumps that corrupt sector aggregates.} For bank $i$ on day $t$, MES is the average loss on bank $i$'s equity over the days in a 252-day rolling window on which the S\&P~500 return fell below its 5th percentile:
\begin{equation}
    \widehat{\mathrm{MES}}_{i,t}^{(0.05)} = \frac{1}{|\mathcal{T}_t|} \sum_{s \in \mathcal{T}_t} (-r_{i,s}), \qquad \mathcal{T}_t = \{s \in W_t : r_{m,s} \leq \widehat{\mathrm{VaR}}_{m,t}^{(0.05)}\},
    \label{c2:eq:mes}
\end{equation}
following \citet{AcharyaEtAl2017}. Bank-level MES is aggregated to a sector index with market-cap weights, resampled to end-of-week frequency and first-differenced.

\subsection{Yield-Curve Factor Extraction}
\label{c2:subsec:pca}

Following \citet{Litterman1991}, principal components are extracted from zero-rate levels rather than changes, because it is the state of the curve that conditions the macro-financial block and because levels are what the mean-reverting factor dynamics describe. Let $y_t = (y_t^{1Y}, y_t^{2Y}, y_t^{5Y}, y_t^{10Y}, y_t^{30Y})'$. Yields are centred and standardised across maturities, and PCA is applied to the standardised level matrix:
\begin{equation}
    \tilde{y}_t = L f_t + \varepsilon_t,
    \label{c2:eq:pca_levels}
\end{equation}
where $L$ is the loading matrix and $f_t = (f_t^{(1)}, f_t^{(2)}, f_t^{(3)})'$ collects the first three components. Loadings are sign-normalised so that PC1 loads positively across the curve (level), PC2 contrasts short and long maturities (slope) and PC3 captures curvature. Table~\ref{c2:tab:pca_loadings} shows that this pattern is recovered and that the three components explain 99.9\% of the variation in zero-rate levels. With five pillars that share is high partly by construction; the substantive point is the shape of the loadings, which reproduces the canonical structure of \citet{Litterman1991} and \citet{DieboldLi2006}. Weekly factor changes $\Delta f_t$ serve as the exogenous inputs to the VARX.

\begin{table}[!ht]
\centering
\caption{Principal Component Loadings and Explained Variance: LSEG USD Zero-Coupon Curves (2010--2024)}
\label{c2:tab:pca_loadings}
\small
\begin{tabular}{lccc}
\toprule
\textbf{Maturity} & \textbf{PC1 (Level)} & \textbf{PC2 (Slope)} & \textbf{PC3 (Curvature)} \\
\midrule
1Y  & 0.453 & $-0.418$ & $-0.604$ \\
2Y  & 0.465 & $-0.382$ & $-0.050$ \\
5Y  & 0.501 & $-0.085$ & 0.635 \\
10Y & 0.470 & 0.359 & 0.244 \\
30Y & 0.327 & 0.737 & $-0.414$ \\
\midrule
Variance explained (\%) & 78.03 & 21.17 & 0.72 \\
Cumulative (\%) & 78.03 & 99.20 & 99.92 \\
\bottomrule
\end{tabular}
\end{table}

\subsection{Hull--White Three-Factor Scenario Generator}
\label{c2:subsec:hull_white}

The scenario generator is a three-factor Gaussian Hull--White model. The short rate is
\begin{equation}
    r_t = x_{1,t} + x_{2,t} + x_{3,t} + \phi(t),
    \label{c2:eq:hw_short_rate}
\end{equation}
where each latent state follows a zero-mean Ornstein--Uhlenbeck process under the risk-neutral measure $\mathbb{Q}$,
\begin{equation}
    dx_{i,t} = -a_i x_{i,t}\,dt + \sigma_i\,dW^{\mathbb{Q},(i)}_t, \qquad i=1,2,3,
    \label{c2:eq:hw_factors}
\end{equation}
with correlated Brownian increments. The deterministic shift $\phi(t)$ makes the model fit the observed initial discount curve exactly. For maturity interval $\tau = T-t$, the zero-coupon bond price is
\begin{equation}
    P(t,T) = A(t,T) \exp\!\left[-\sum_{i=1}^{3} B_i(t,T)\, x_{i,t}\right],
    \qquad
    B_i(t,T) = \frac{1-e^{-a_i\tau}}{a_i},
    \label{c2:eq:g3pp_bond_price}
\end{equation}
where $A(t,T)$ is computed from the observed discount curve and the model variance adjustment.

\paragraph{Two measures, one model.} Equations~\eqref{c2:eq:hw_factors} and \eqref{c2:eq:g3pp_bond_price} are the pricing block. They give the value of a bond at time $t$ for a given state $x_t$, and they are arbitrage-free because $A(t,T)$ reproduces the observed discount function at $t=0$. They say nothing about how the state evolves in the real world, which is what a scenario generator has to describe. The two questions are linked by the market price of risk. Taking $\lambda_i$ constant, so that $dW^{\mathbb{Q},(i)}_t = dW^{\mathbb{P},(i)}_t + \lambda_i\,dt$, the same factors under the physical measure $\mathbb{P}$ follow
\begin{equation}
    dx_{i,t} = a_i\big(\theta_i - x_{i,t}\big)\,dt + \sigma_i\,dW^{\mathbb{P},(i)}_t,
    \qquad
    \theta_i = \frac{\sigma_i \lambda_i}{a_i},
    \label{c2:eq:p_dynamics}
\end{equation}
which mean-revert to a non-zero level $\theta_i$. The whole term premium of the model is carried by $\theta$: a factor with $\theta_i = 0$ earns no compensation for its risk, and the sign of $\theta_i$ says whether that compensation is positive or negative.

States are simulated forward under \eqref{c2:eq:p_dynamics}, because that is how yields move, and each simulated state is priced through \eqref{c2:eq:g3pp_bond_price}, because that is what the resulting curve is worth. A generator that simulated under $\mathbb{Q}$ instead would produce curves drifting at the risk-free rate and would understate the persistence of observed yield levels.

\paragraph{Physical-measure calibration.} At each estimation date the LSEG panel is decomposed into three principal components, mean-reversion speeds are estimated from the factor dynamics and ordered from fast to slow, and historical zero-rate deviations are projected onto the maturity-sensitivity basis $B_i(\tau)/\tau$ to produce a time series of latent states. State dynamics are estimated by the exact discretisation of \eqref{c2:eq:p_dynamics},
\begin{equation}
    x_{i,t+\Delta t} = \theta_i + e^{-a_i\Delta t}(x_{i,t}-\theta_i)
    + \sigma_i \sqrt{\frac{1-e^{-2a_i\Delta t}}{2a_i}}\,\eta_{i,t},
    \label{c2:eq:ou_discretisation}
\end{equation}
where the innovation vector $\eta_t$ has empirical correlation matrix $R$, regularised to be positive definite. A change of measure alters the drift but not the diffusion, so $a_i$, $\sigma_i$ and $R$ are common to both measures and only $\theta$ is measure-specific. Estimating $\theta$ from history is therefore the same as estimating the market price of risk. Table~\ref{c2:tab:lambda_implied} reports the implied $\lambda_i = a_i\theta_i/\sigma_i$ at the December 2024 anchor. None exceeds 0.09 in absolute value, so the fitted physical drift corresponds to a modest and economically ordinary risk premium.

\begin{table}[!ht]
\centering
\caption{Fitted Factor Parameters and the Implied Market Price of Risk}
\label{c2:tab:lambda_implied}
\small
\begin{tabular}{lcccc}
\toprule
Factor & $a_i$ & $\theta_i$ & $\sigma_i$ & Implied $\lambda_i$ \\
\midrule
$x_1$ (fast)   & 1.557 & $-0.00210$ & 0.0498 & $-0.066$ \\
$x_2$ (medium) & 0.460 & $+0.01342$ & 0.0710 & $+0.087$ \\
$x_3$ (slow)   & 0.140 & $-0.00599$ & 0.0392 & $-0.021$ \\
\bottomrule
\end{tabular}
\begin{flushleft}
\footnotesize Estimated on the full sample at the December 2024 anchor. $\lambda_i = a_i\theta_i/\sigma_i$ from \eqref{c2:eq:p_dynamics}. Simulation is initialised at $x_0 = 0$, where the model reproduces the observed discount curve exactly.
\end{flushleft}
\end{table}

\paragraph{A constant price of risk.} Holding $\lambda_i$ constant places the model in the completely affine class of \citet{Duffee2002}, which ties the compensation for risk to factor volatility and cannot reproduce time variation in expected excess bond returns; in Duffee's sample such models did not forecast yield levels better than a random walk. His essentially affine class lets the price of risk depend on the state and is the natural extension. The tighter specification is kept here for two reasons. The block is used to generate a distribution of plausible curves, not to extract the most accurate conditional mean, and a state-dependent price of risk adds parameters estimated from the same fifteen years of data that already identify the factor dynamics. A constant $\lambda$ keeps the term premium as one inspectable number per factor. And the out-of-sample results in Section~\ref{c2:subsec:results_phase7} are consistent with what Duffee's finding predicts: the block has lower RMSE than the random walk in most cells, but not significantly so.

\paragraph{Volatility calibration and curve fit.} State volatilities are scaled so that the model's analytic weekly zero-rate volatility matches the empirical weekly volatility of the LSEG panel. The model is simulated at a daily step and sampled at weekly Friday horizons, and each simulated curve is projected through the training-sample PCA map to give simulated weekly $\Delta$PC paths for the VARX. The initial curve is fitted exactly: at the December 2024 anchor the maximum zero-rate fit error across the five pillars is below $10^{-12}$\,bp, so forecast errors come from the physical dynamics, not from the initial fit. Table~\ref{c2:tab:g3pp_vol_fit} shows that the model matches the 1Y--10Y volatility profile closely (simulated-to-empirical ratios of 1.03 to 1.19) and underfits the 30Y pillar (0.67). The 30Y pillar is retained as a diagnostic, since the policy-relevant transmission horizon lies in the 1Y--10Y segment.

\begin{table}[!ht]
\centering
\caption{G3++ Weekly Zero-Rate Volatility Fit}
\label{c2:tab:g3pp_vol_fit}
\small
\begin{tabular}{lcccc}
\toprule
Maturity & Empirical vol. & Analytic G3++ vol. & Simulated G3++ vol. & Sim./Emp. \\
& (bp/week) & (bp/week) & (bp/week) & ratio \\
\midrule
1Y  & 7.58 & 8.59 & 8.85 & 1.17 \\
2Y  & 9.38 & 9.50 & 9.67 & 1.03 \\
5Y  & 11.01 & 11.98 & 12.14 & 1.10 \\
10Y & 11.34 & 13.28 & 13.46 & 1.19 \\
30Y & 10.90 & 7.24 & 7.34 & 0.67 \\
\bottomrule
\end{tabular}
\end{table}

\subsection{VARX Estimation with Minnesota-Prior Shrinkage}
\label{c2:subsec:varx}

The macro-financial block is a VARX(2) with three exogenous yield-curve inputs:
\begin{equation}
Y_t = c + A_1Y_{t-1} + A_2Y_{t-2} + B_x X_t + u_t,
\qquad
u_t \sim \mathcal{N}(0,\Sigma_u),
\end{equation}
where
\begin{gather*}
Y_t = (\texttt{mkt\_ret}_t,\ \texttt{dffr}_t,\ \texttt{dwei}_t,\ \texttt{dinf}_t,\ \texttt{d\_mes\_proxy}_t)',\\
X_t = (\Delta PC1_t,\ \Delta PC2_t,\ \Delta PC3_t)'.
\end{gather*} The lag order is fixed at $p=2$; longer lag structures are poorly identified in a five-variable weekly system. The yield-curve factors are exogenous rather than endogenous, which reflects the separation between G3++ as the upstream generator and the VARX as the downstream transmission layer; modelling them endogenously would reintroduce feedback from macro variables to yields and break that separation.

Estimation uses the Minnesota prior of \citet{Doan1984} and \citet{Litterman1986}, which shrinks each lag coefficient towards zero with a variance that shrinks faster for longer lags and for cross-variable terms. For an endogenous lag coefficient $\beta_{i,j,l}$,
\begin{equation}
\beta_{i,j,l} \sim \mathcal{N}\!\left(
0,\;
\lambda^2 \frac{s_i^2}{s_j^2} \frac{1}{l^{2d}} \times
\begin{cases}
1, & i=j,\\
\lambda_{\mathrm{cross}}^2, & i\neq j,
\end{cases}
\right),
\end{equation}
where $s_i, s_j$ are sample standard deviations, $\lambda$ is overall tightness, $\lambda_{\mathrm{cross}}$ controls cross-variable shrinkage and $d$ governs lag decay. The prior mean is zero for all lag coefficients because the variables are returns and first differences. Intercept and exogenous-coefficient priors are looser, controlled by $\lambda_{\mathrm{const}}$ and $\lambda_{\mathrm{exog}}$. Baseline hyperparameters are $(\lambda, \lambda_{\mathrm{cross}}, d, \lambda_{\mathrm{const}}, \lambda_{\mathrm{exog}}) = (0.5, 0.8, 1.0, 10.0, 5.0)$; sensitivity to $\lambda$ is reported in Section~\ref{c2:sec:minnesota_sensitivity}. The estimated system is stationary, with a maximum companion-matrix eigenvalue modulus of 0.593. The estimation sample has 757 weekly observations from 2 July 2010 to 27 December 2024.

\subsection{Structural Identification}
\label{c2:subsec:identification}

A contractionary monetary-policy shock is identified by sign restrictions \citep{Uhlig2005, RubioRamirez2010}, which do not pin down a single shock but a set of admissible ones, augmented by a relevance filter on the policy-rate loading and by magnitude caps.

\subsubsection{Sign Restrictions}
\label{c2:subsubsec:sign_restrictions}

The reduced-form covariance matrix is factorised as $\Sigma_u = CC'$ by Cholesky decomposition, and candidate impact matrices $B = CQ$ are generated by drawing orthonormal matrices $Q$ uniformly at random. A rotation is retained only if
\begin{enumerate}[nosep]
    \item $\texttt{dffr}(h=0) > 0$: the policy rate rises on impact;
    \item $\texttt{mkt\_ret}(h) < 0$ for $h = 0, 1, 2$: equity returns fall on impact and stay below baseline for two further weeks.
\end{enumerate}
The equity restriction covers three weeks because a one-week restriction admits rotations in which the equity response reverses immediately, which is not the mechanism being identified. The responses of the WEI, inflation and MES are left unrestricted so that the data can speak on the systemic-risk channel.

\subsubsection{Relevance Filter and Magnitude Caps}
\label{c2:subsubsec:relevance_filter}

In weekly monetary VARs the standard approach has a specific failure mode. Rotations that place almost no weight on the policy-rate dimension can satisfy $\texttt{dffr}(h=0) > 0$ while carrying essentially no policy-rate signal. When each accepted rotation is then rescaled to a fixed impact, here $+25$\,bp, dividing by that small loading inflates every other response, producing implausibly large magnitudes and the price-puzzle pattern documented by \citet{Sims1992}. The problem is acute at weekly frequency because the policy rate is constant in most weeks, so the residual variance of the funds-rate equation is small relative to those of equity returns and the WEI.

The \emph{relevance filter} requires each candidate rotation to place a meaningful loading on the policy rate before normalisation:
\begin{equation}
    |v_0[\texttt{dffr}]| \geq \tau \cdot \sigma_{\texttt{dffr}}, \qquad \tau = 0.20,
    \label{c2:eq:relevance_filter}
\end{equation}
where $\sigma_{\texttt{dffr}}$ is the residual standard deviation of the funds-rate equation and $v_0$ is the impact vector in standardised units. Table~\ref{c2:tab:tau_sensitivity} reports a diagnostic sweep over $\tau \in \{0.10,\allowbreak\ 0.20,\allowbreak\ 0.30,\allowbreak\ 0.40,\allowbreak\ 0.50\}$ on a fixed budget of 500,000 draws. The number of sign-admissible columns, 1,431, is the same at every $\tau$ by construction. The share passing the relevance filter falls from 23\% at $\tau=0.10$ to 1.2\% at $\tau=0.20$ and to nothing at $\tau=0.30$ or above, so $\tau=0.20$ is the largest threshold that leaves a non-empty set and is the value used. At that budget a single rotation also passes the magnitude caps, which is why the final search below runs to five million draws.

\begin{table}[!ht]
\centering
\caption{Relevance-Filter Sensitivity: Diagnostic Sweep on 500,000 Draws}
\label{c2:tab:tau_sensitivity}
\small
\begin{tabular}{lcccc}
\toprule
$\tau$ & Sign-admissible & Relevant & Relevant share (\%) & Pass magnitude caps \\
\midrule
0.10 & 1,431 & 323 & 22.6 & 1 \\
0.20 & 1,431 & 17  & 1.2  & 1 \\
0.30 & 1,431 & 0   & 0.0  & 0 \\
0.40 & 1,431 & 0   & 0.0  & 0 \\
0.50 & 1,431 & 0   & 0.0  & 0 \\
\bottomrule
\end{tabular}
\begin{flushleft}
\footnotesize Fixed budget of 500,000 Haar draws, all five columns of each screened, magnitude caps as in Table~\ref{c2:tab:sign_restrictions}. The final adaptive search reported below runs for five million draws.
\end{flushleft}
\end{table}

\emph{Magnitude caps} are applied to three impact responses: the sign-restricted equity response and the two unrestricted real-side responses. They are fixed at $|\Delta\texttt{mkt\_ret}(h=0)| \leq 0.075$, $|\Delta\texttt{dwei}(h=0)| \leq 0.045$ and $|\Delta\texttt{dinf}(h=0)| \leq 0.012$, values calibrated during model development at the 75th percentile of the absolute impact response across the rotations that survived the relevance filter, and held fixed thereafter. They remove rotations that pass the sign and relevance criteria but imply economically implausible magnitudes, such as a weekly equity decline above 7.5\% from a single 25\,bp surprise. Two properties of the caps should be stated plainly. On the final estimation the relevant set differs from the development sample, so the caps no longer sit at a common percentile: the equity cap does not bind, the WEI cap sits near the 80th percentile of the relevant set, and the inflation cap near the 25th, so the three together retain 41 of 300 relevant rotations. And they bind near the centre of the admissible distribution rather than in its tail: the accepted medians in Section~\ref{c2:subsec:results_irf} are 91\% and 90\% of the WEI and inflation caps. The identified set is therefore the lower portion of a distribution whose upper portion was removed as implausible, which bears on how the magnitudes are read. Section~\ref{c2:subsec:results_irf} shows that the sign pattern is the same with no caps, with caps recalibrated by the same percentile rule on the final sample, and with the fixed caps, so the caps shape the scale of the responses but not their direction.

The adaptive search was run for five million draws with all five columns of each screened: 14,779 columns satisfied the sign restrictions, 300 of those passed the relevance filter and 41 passed the caps. Impulse-response bands are the 10th, 50th and 90th percentiles across the 41 accepted rotations, so inference is set-identified rather than based on a single rotation.

\begin{table}[!ht]
\centering
\caption{Sign Restrictions, Relevance Filter, and Magnitude Caps for the Contractionary MP Shock}
\label{c2:tab:sign_restrictions}
\small
\begin{tabular}{lcc}
\toprule
\textbf{Variable / Filter} & \textbf{Sign (h=0)} & \textbf{Magnitude Cap} \\
\midrule
FFR change (\texttt{dffr}) & $+$ & --- \\
FFR loading (relevance filter) & --- & $|v_0[\texttt{dffr}]| \geq 0.20\,\sigma_{\texttt{dffr}}$ \\
Market return (\texttt{mkt\_ret}) & $-$ ($h=0,1,2$) & $\leq 7.5\%$ weekly \\
WEI change (\texttt{dwei}) & unrestricted & $\leq 4.5$ pp \\
Inflation change (\texttt{dinf}) & unrestricted & $\leq 1.2$ pp \\
MES proxy (\texttt{d\_mes\_proxy}) & unrestricted & --- \\
\bottomrule
\end{tabular}
\end{table}

\subsubsection{What the Exogenous Block Implies for the Identified Shock}
\label{c2:subsubsec:orthogonality}

Because the yield-curve factor changes enter the VARX contemporaneously as exogenous regressors, the reduced-form innovations $u_t$ are orthogonal to the same week's movement in level, slope and curvature, and so is any structural shock recovered from them. The identified policy shock is therefore the part of the weekly funds-rate innovation that is not spanned by that week's curve change. A monetary surprise ordinarily moves the curve on impact, so part of what a textbook policy shock does is absorbed by the conditioning variables rather than attributed to the shock. This is one further reason the impact responses in Section~\ref{c2:subsec:results_irf} are read for sign and relative size rather than for scale, and it is why the identification is also run with the exogenous block switched off. Section~\ref{c2:subsec:results_irf} reports that check: the sign pattern is unchanged, and what changes is how hard the shock is to isolate.

\subsection{Conditional Forecasting and Stress Scenarios}
\label{c2:subsec:conditional_forecast}

The pipeline produces conditional forecasts by feeding G3++-simulated weekly $\Delta$PC paths into the VARX. For each simulation $s \in \{1,\dots,S\}$ and horizon $h$,
\begin{equation}
    \hat{Y}_{t+h}^{(s)} = c + \sum_{l=1}^{p} A_l \hat{Y}_{t+h-l}^{(s)} + B_x X_{t+h}^{(s)} + u_{t+h}^{(s)},
    \qquad u_{t+h}^{(s)} \sim \mathcal{N}(0, \Sigma_u),
    \label{c2:eq:conditional_forecast}
\end{equation}
where $X_{t+h}^{(s)}$ is the $s$-th simulated path. The cross-section of $\hat{Y}_{t+h}^{(s)}$ gives forecast medians and quantile bands. For stress scenarios, the generator is initialised from a shocked anchor curve, such as a parallel $+200$\,bp shift, an inversion or a bull steepener, and the OU dynamics propagate the shock while the VARX translates it into macro-financial outcomes.

A scenario in this framework is a path of curve movements, not a level. The transmission layer conditions on weekly factor changes, so a curve that is simply placed at a stressed level and left there conveys nothing to it: with the OU dynamics started from the shocked curve, the simulated factor changes are the same as from the unshocked one, and the three level scenarios were verified to produce distributions that differ only by simulation noise. The scenarios are therefore specified as ramps. The curve moves linearly from the December 2024 anchor to the shocked curve over 26 weeks and stays there, the G3++ dynamics run around that path, and a no-shock baseline is simulated with the same random numbers so that each scenario is read as a difference from it. For the parallel $+200$\,bp case the ramp is 7.7\,bp a week at every pillar, about one historical weekly standard deviation and roughly the pace of the 2022 tightening cycle, so the VARX sees factor changes inside the range it was estimated on. The earlier implementation, which entered the whole shift as a single-week factor change many standard deviations wide, is the artefact this design replaces.

\subsection{Out-of-Sample Evaluation Design}
\label{c2:subsec:oos_design}

Evaluation uses an expanding window with origins 13 weeks apart. At each origin every component, the PCA loadings, the G3++ parameters and the VARX coefficients, is re-estimated on data up to that date only. Three tiers are run:

\begin{enumerate}[nosep,leftmargin=*]
    \item \textbf{Yield-curve forecasts.} G3++ yield forecasts against a random walk and a direct VAR(1) on weekly zero-rate changes, with RMSE in basis points at $h \in \{4,12,26,52\}$ weeks and Diebold--Mariano tests of equal accuracy with HAC standard errors.
    \item \textbf{VARX endogenous block.} Two modes: \emph{Mode A} conditions on the realised future PC changes, which isolates transmission skill from scenario error; \emph{Mode B} sets the exogenous PC changes to zero, which tests the unconditional dynamics. Benchmarks are an AR(1) and a no-change (Zero) forecast.
    \item \textbf{Joint pipeline.} G3++-simulated PC paths feed the re-estimated VARX. Point accuracy is measured by RMSE of the forecast median. Density accuracy is measured by probability integral transform (PIT) histograms, which are flat when the forecast distribution is correctly calibrated; by Kupiec coverage tests of 80\% intervals; and by the continuous ranked probability score (CRPS), which scores the whole predictive distribution against the outcome and can be compared across models.
\end{enumerate}

The yield-curve evaluation needs 52 weeks of realised data after each origin, giving 29 origins ending December 2023; the VARX and joint-pipeline evaluations need 26 weeks, giving 31 origins ending June 2024. All begin in January 2017.

One consequence of the fixed 13-week spacing recurs in the tables. Every 26-week target date coincides with a later origin week, and every 12-week target date falls one week before one, so those cells each sample a fixed week of the quarter across all 31 origins. For the two intermittently moving series this produces cells in which the realised weekly change is zero at every origin: no FOMC rate change fell in a 26-week target week and no CPI release fell in a 12-week target week. In those two cells (\texttt{dffr} at 26 weeks and \texttt{dinf} at 12 weeks) the Zero benchmark has an RMSE of exactly zero and any interval covers the outcome at every origin, so any other forecast loses to Zero by construction. They are excluded from all Diebold--Mariano statements below.

\section{Results and Discussion}
\label{c2:sec:results}

\phantomsection\label{c2:subsec:results_irf}
The structural block comes first, because its impact responses are used later to read the stress scenarios. %

\begin{figure}[!ht]
\centering
\includegraphics[width=0.9\textwidth]{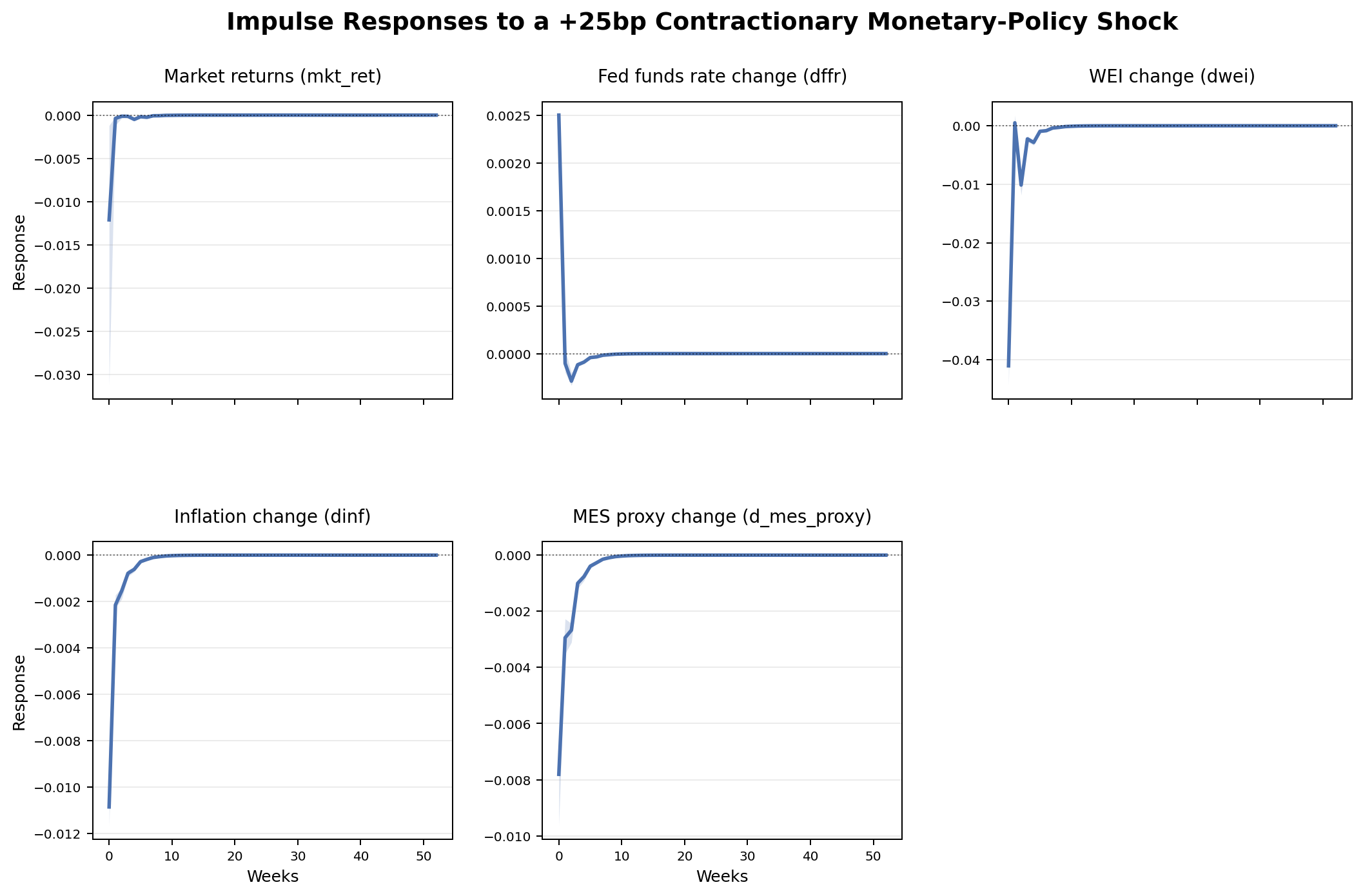}
\caption{Impulse responses to a $+25$\,bp contractionary policy shock, first twelve weeks. Lines: medians across accepted rotations; shading and the bar at impact: 10--90\% set-identified bands.}
\label{c2:fig:irf_macro}
\end{figure}

Table~\ref{c2:tab:phase2_impact_responses} reports impact responses to the identified contractionary shock, normalised to $+25$\,bp, and Figure~\ref{c2:fig:irf_macro} shows the dynamic paths. Bands span the 41 accepted rotations and are a plausibility-restricted identified set rather than the full set of sign-admissible rotations.

The policy-rate change rises on impact by construction and decays quickly, as expected for a variable measured in weekly changes. Equity returns fall on impact (median $-1.21\%$). Activity and prices respond contractionarily: the median WEI response is $-4.11$\,pp and the median CPI response is $-1.08$\,pp.

These magnitudes are large relative to the variables' own weekly variation and should not be read as point estimates of the effect of a 25\,bp surprise. Three features of the design produce them. Normalising each accepted rotation to a $+25$\,bp impact divides every other response by that rotation's policy-rate loading, which is small at weekly frequency even after the relevance filter. The magnitude caps bind near the centre of the admissible distribution, so the identified set is the lower portion of a distribution whose upper portion was removed. And the shock is orthogonal to the same week's curve movement (Section~\ref{c2:subsubsec:orthogonality}). What survives these features is the sign pattern and the comparison across indicators, and that is how the responses are used: to establish the direction of the activity, price and tail-risk responses, and to contrast MES with the covariance proxy. A design intended to quantify the size of the macroeconomic effect of a policy surprise would need a lower-frequency policy variable or an external-instrument identification, neither of which is available at weekly frequency for the reasons given in Section~\ref{c2:sec:literature}.

The MES proxy falls on impact (median $-0.78\%$, 80\% band $[-0.96\%, -0.68\%]$). This does not mean that tightening reduces bank-sector fragility. MES is a level measure based on equity returns, the average loss on bank equity on market-tail days, and at weekly frequency its immediate response is dominated by the discount-rate channel: a tightening surprise lowers all equity prices, including bank stocks.

\begin{table}[!ht]
\centering
\caption{Impact Responses ($h=0$) to a $+25$\,bp Contractionary MP Shock}
\label{c2:tab:phase2_impact_responses}
\small
\begin{tabular}{lccc l}
\toprule
Variable & p10 & Median & p90 & Interpretation \\
\midrule
\texttt{mkt\_ret}      & $-3.13$  & $-1.21$  & $-0.12$  & Weekly return, percent \\
\texttt{dffr}          & $+25.00$ & $+25.00$ & $+25.00$ & Weekly $\Delta$FFR, bp \\
\texttt{dwei}          & $-4.44$  & $-4.11$  & $-3.38$  & Weekly $\Delta$WEI, pp \\
\texttt{dinf}          & $-1.16$  & $-1.08$  & $-0.91$  & Weekly $\Delta$CPI YoY, pp \\
\texttt{d\_mes\_proxy} & $-0.96$  & $-0.78$  & $-0.68$  & Weekly $\Delta$MES proxy, percent \\
\bottomrule
\end{tabular}
\begin{flushleft}
\footnotesize Percentiles across 41 accepted rotations satisfying the sign, relevance and magnitude criteria. Shock normalised so $\texttt{dffr}(h=0)=+25$\,bp. Search over 5{,}000{,}000 rotations, screening all five columns of each: 14{,}779 columns passed the sign restrictions, 300 passed the relevance filter, 41 passed the magnitude caps.
\end{flushleft}
\end{table}

To clarify what the MES response measures, the system is re-estimated with a bank--market covariance proxy in place of $\Delta$MES, holding all identification settings fixed. The proxy is a rolling covariance between winsorised bank and market daily returns, market-cap weighted to the sector, resampled to weekly frequency and first-differenced. It is inspired by CoVaR \citep{AdrianBrunnermeier2016} in that it targets co-movement between bank-sector and market outcomes, but it is not the quantile-regression CoVaR of that paper, and it is labelled the covariance proxy throughout to keep the distinction clear. Across 100 accepted rotations its median impact response is $+1.54\times 10^{-5}$ in decimal-squared units, the units of a covariance of decimal returns, with the entire 80\% band positive, $[+4.55\times 10^{-6}, +2.73\times 10^{-5}]$. The covariance-proxy system reaches 100 accepted rotations in 36,930 draws, against 41 in five million for the MES system, because replacing the systemic variable changes the residual covariance and with it the share of rotations that place weight on the policy rate.

The two indicators therefore move in opposite directions under the same shock, and the divergence is informative. MES is a level measure pushed down mechanically by the discount-rate channel. The covariance proxy is a co-movement measure, and its rise under tightening is consistent with the risk-taking channel of \citet{BorioZhu2012}: tightening exposes accumulated vulnerabilities and increases bank--market co-movement even as the level of bank equity falls. MES is best read as a tail-risk pricing indicator and the covariance proxy as a co-movement indicator closer to the structural risk-taking interpretation. A full CoVaR implementation would sharpen the second and is listed among the extensions in Section~\ref{c2:sec:conclusion}.

One more check on the identification is needed before the forecast evaluation. Section~\ref{c2:subsubsec:orthogonality} noted that the identified shock is orthogonal to the same week's curve movement by construction. To check that this does not drive the sign pattern, the VARX is re-estimated with the exogenous block switched off and the rotation search repeated with the same seed, threshold and horizon restrictions, collecting 150 relevant rotations in each specification. Table~\ref{c2:tab:modeB} reports the impact bands on the relevant set with no caps, so that the comparison does not depend on the caps at all, and with caps recalibrated by the 75th-percentile rule on each specification's own relevant set.

Two things stand out. The sign pattern survives: every median response has the same sign with and without the curve, and the equity, inflation and MES bands lie entirely below zero in both. The WEI response is the one that weakens without the curve, with a median of $-0.99$\,pp against $-3.06$\,pp and an 80\% band whose upper end just crosses zero, so activity is where the conditioning variables carry most of the shock's content. And the shock becomes much harder to isolate: the share of sign-admissible rotations that pass the relevance filter falls from 2.1\% with the curve to 0.3\% without it, so collecting the same 150 relevant rotations takes 14.7 million draws instead of 2.4 million, and none of the rotations found without the curve satisfies the fixed caps of Table~\ref{c2:tab:sign_restrictions}. Without the curve, the funds-rate innovation is too small a component of the weekly system to carry a 25\,bp normalisation, which is the practical reason the exogenous block is needed for identification at this frequency. The same exercise on the covariance-proxy system gives near-identical bands with and without the curve, and the proxy's response stays positive in both.

\begin{table}[!ht]
\centering
\caption{Impact Responses With and Without the Yield-Curve Block}
\label{c2:tab:modeB}
\small
\resizebox{\textwidth}{!}{%
\begin{tabular}{l ccc ccc ccc ccc}
\toprule
& \multicolumn{6}{c}{Relevant set, no caps ($n=150$ each)} & \multicolumn{6}{c}{Same-rule caps ($n=53$ / $n=49$)} \\
\cmidrule(lr){2-7}\cmidrule(lr){8-13}
& \multicolumn{3}{c}{With curve} & \multicolumn{3}{c}{Without curve} & \multicolumn{3}{c}{With curve} & \multicolumn{3}{c}{Without curve} \\
\cmidrule(lr){2-4}\cmidrule(lr){5-7}\cmidrule(lr){8-10}\cmidrule(lr){11-13}
Variable & p10 & p50 & p90 & p10 & p50 & p90 & p10 & p50 & p90 & p10 & p50 & p90 \\
\midrule
\texttt{mkt\_ret} (\%)      & $-3.26$ & $-1.20$ & $-0.26$ & $-5.03$ & $-2.24$ & $-0.33$ & $-1.86$ & $-0.81$ & $-0.24$ & $-3.26$ & $-1.29$ & $-0.37$ \\
\texttt{dwei} (pp)          & $-4.75$ & $-3.06$ & $-1.31$ & $-2.39$ & $-0.99$ & $+0.18$ & $-3.76$ & $-3.00$ & $-2.17$ & $-1.49$ & $-0.73$ & $+0.11$ \\
\texttt{dinf} (pp)          & $-1.91$ & $-1.42$ & $-0.91$ & $-2.22$ & $-2.08$ & $-1.91$ & $-1.63$ & $-1.41$ & $-1.12$ & $-2.14$ & $-2.07$ & $-1.99$ \\
\texttt{d\_mes\_proxy} (\%) & $-1.21$ & $-0.99$ & $-0.71$ & $-1.13$ & $-0.96$ & $-0.82$ & $-1.08$ & $-0.94$ & $-0.77$ & $-1.03$ & $-0.87$ & $-0.79$ \\
\bottomrule
\end{tabular}}
\begin{flushleft}
\footnotesize Impact responses ($h=0$) to a $+25$\,bp shock, same seed, $\tau=0.20$ and $H_{\mathrm{sign}}=2$ as Table~\ref{c2:tab:phase2_impact_responses}. ``Without curve'' sets the exogenous block to zero (Mode~B of Section~\ref{c2:subsec:oos_design}). Same-rule caps are the 75th percentile of $|h=0|$ on each specification's own relevant set (with curve: 2.3\%, 4.1\,pp, 1.8\,pp; without: 3.8\%, 1.7\,pp, 2.2\,pp). Relevant share of sign-admissible rotations: 2.1\% with the curve, 0.3\% without. Share of the WEI response below zero without the curve: 85\%.
\end{flushleft}
\end{table}

\FloatBarrier

\phantomsection\label{c2:subsec:results_phase7}
The out-of-sample evaluation begins with the term-structure block on its own. %

Table~\ref{c2:tab:phase7_rmse} reports out-of-sample RMSE in basis points for G3++, the direct VAR(1) and the random walk across 29 origins. G3++ has the lowest RMSE outright in 13 of 20 maturity--horizon cells and ties the random walk in a fourteenth (10Y at $h=26$), with the strongest performance at the 1Y, 2Y and 5Y pillars. Diebold--Mariano tests, with HAC bandwidths capped by the number of origins, do not reject equal accuracy against the random walk at the 5\% level in any cell. The loss differential favours G3++ at the front and belly of the curve and the random walk at parts of the long end.

The block is therefore a competitive curve generator rather than a dominant conditional-mean forecaster. That is the outcome its constant price of risk predicts: \citet{Duffee2002} shows that completely affine models should not be expected to beat a random walk in level forecasts, and the pattern here, lower RMSE in most cells without significant rejections, is what that implies. It is also the property the block needs for its role, which is to simulate arbitrage-consistent distributions of future curves rather than to forecast their mean.

\begin{table}[!ht]
\centering
\caption{Yield-Curve Out-of-Sample RMSE by Horizon and Maturity (bp)}
\label{c2:tab:phase7_rmse}
\scriptsize
\resizebox{\textwidth}{!}{%
\begin{tabular}{l ccc ccc ccc ccc}
\toprule
& \multicolumn{3}{c}{$h=4$w} & \multicolumn{3}{c}{$h=12$w} & \multicolumn{3}{c}{$h=26$w} & \multicolumn{3}{c}{$h=52$w} \\
\cmidrule(lr){2-4}\cmidrule(lr){5-7}\cmidrule(lr){8-10}\cmidrule(lr){11-13}
Mat. & G3++ & VAR & RW & G3++ & VAR & RW & G3++ & VAR & RW & G3++ & VAR & RW \\
\midrule
1Y  & 12.8 & 16.0 & 16.3 & 46.0 & 53.5 & 54.3 & 82.7 & 96.9 & 97.7 & 151.8 & 177.5 & 173.9 \\
2Y  & 14.1 & 16.0 & 15.8 & 52.4 & 55.1 & 55.7 & 83.8 & 90.7 & 91.1 & 144.1 & 161.7 & 157.7 \\
5Y  & 18.6 & 19.0 & 18.9 & 49.4 & 49.6 & 50.2 & 74.4 & 76.4 & 76.6 & 121.1 & 129.5 & 125.2 \\
10Y & 21.5 & 21.7 & 21.5 & 47.1 & 46.8 & 47.1 & 68.8 & 69.4 & 68.8 & 107.6 & 111.8 & 106.8 \\
30Y & 20.3 & 20.5 & 20.2 & 43.1 & 42.8 & 42.9 & 59.4 & 60.0 & 58.6 & 91.1 & 95.2 & 89.2 \\
\bottomrule
\end{tabular}}
\begin{flushleft}
\footnotesize G3++ has the lowest RMSE outright in 13 of 20 cells and ties the random walk at 10Y, $h=26$. At two decimals G3++ is also lower at 10Y, $h=4$ (21.47 against 21.54).
\end{flushleft}
\end{table}

\FloatBarrier

\phantomsection\label{c2:subsec:results_phase8}
The transmission layer is evaluated next, before the two are joined. %

Table~\ref{c2:tab:phase8_rmse} reports the endogenous-block RMSE in both conditioning modes alongside the AR(1) and Zero benchmarks. At $h=4$, conditioning on the realised PC path (Mode~A) gives the best forecast for the MES proxy and the funds-rate change, while Zero or the AR(1) is best for the other three variables. At $h=26$ a simple benchmark is best in every cell, and Diebold--Mariano tests find Mode~A significantly worse than Zero for \texttt{mkt\_ret}, \texttt{dwei}, \texttt{dinf} and \texttt{d\_mes\_proxy}. Realised yield-curve paths therefore add little point-forecast accuracy for weekly first-differenced variables once their own lags are included. The role of the VARX is scenario transmission, mapping a given curve path into a coherent macro-financial distribution, rather than point-forecast dominance.

\begin{table}[!ht]
\centering
\caption{VARX Endogenous Block OOS RMSE (Selected Horizons)}
\label{c2:tab:phase8_rmse}
\small
\resizebox{\textwidth}{!}{%
\begin{tabular}{l cccc cccc}
\toprule
& \multicolumn{4}{c}{$h=4$ weeks} & \multicolumn{4}{c}{$h=26$ weeks} \\
\cmidrule(lr){2-5}\cmidrule(lr){6-9}
Variable & VARX$_A$ & VARX$_B$ & AR(1) & Zero & VARX$_A$ & VARX$_B$ & AR(1) & Zero \\
\midrule
\texttt{mkt\_ret}      & 0.023452 & 0.022521 & 0.022581 & 0.022477 & 0.019630 & 0.016634 & 0.016614 & 0.017146 \\
\texttt{dffr}          & 0.001541 & 0.001558 & 0.001553 & 0.001555 & 0.000158 & 0.000043 & 0.000043 & 0.000000 \\
\texttt{dwei}          & 0.002529 & 0.002519 & 0.002471 & 0.002458 & 0.005848 & 0.005774 & 0.005774 & 0.005772 \\
\texttt{dinf}          & 0.001544 & 0.001527 & 0.001526 & 0.001526 & 0.004161 & 0.004130 & 0.004131 & 0.004127 \\
\texttt{d\_mes\_proxy} & 0.001180 & 0.001233 & 0.001249 & 0.001254 & 0.001046 & 0.000994 & 0.000993 & 0.000974 \\
\bottomrule
\end{tabular}}
\begin{flushleft}
\footnotesize The Zero RMSE of exactly 0.000000 for \texttt{dffr} at $h=26$ reflects a realised change of zero at every origin; see Section~\ref{c2:subsec:oos_design}.
\end{flushleft}
\end{table}

\FloatBarrier

\phantomsection\label{c2:sec:minnesota_sensitivity}

A natural concern is that the rapid decay of the impulse responses is imposed by the prior. Table~\ref{c2:tab:minnesota_sensitivity} re-estimates the standardised VARX over $\lambda \in \{0.2, 0.5, 1.0, 2.0\}$ with the other prior settings fixed. The maximum companion root stays below 0.60 across the grid, the implied half-life moves only from 1.28 to 1.34 weeks, and residual RMSE is unchanged. The fast decay reflects the low persistence of weekly first-differenced variables, not the prior.

\begin{table}[!ht]
\centering
\caption{VARX Sensitivity to Minnesota-Prior Tightness}
\label{c2:tab:minnesota_sensitivity}
\small
\begin{tabular}{lccccc}
\toprule
$\lambda$ & Lag order & Max root & Half-life & Avg.\ std.\ RMSE & $\sum_l \|A_l\|_F$ \\
\midrule
0.2 & 2 & 0.581 & 1.28 & 0.933 & 0.823 \\
0.5 & 2 & 0.593 & 1.33 & 0.933 & 0.871 \\
1.0 & 2 & 0.595 & 1.34 & 0.933 & 0.879 \\
2.0 & 2 & 0.596 & 1.34 & 0.933 & 0.881 \\
\bottomrule
\end{tabular}
\end{table}

\FloatBarrier

\phantomsection\label{c2:subsec:results_phase9a}
With both blocks evaluated separately, the pipeline can be evaluated as a whole. %

Table~\ref{c2:tab:phase9_rmse} evaluates the full pipeline: at each origin the PCA map, the G3++ dynamics and the VARX are re-estimated, and G3++ generates the PC paths that condition the VARX. The pipeline has lower RMSE than Zero in 8 of 15 variable--horizon cells, with the clearest gains for market returns (12w, 26w), the MES proxy (all horizons) and the funds-rate change (4w, 12w). Diebold--Mariano tests reject equal accuracy in two cells, in opposite directions: for market returns at 26 weeks in the pipeline's favour, and for inflation at 26 weeks against it.

The headline count understates the comparison. Six of the fifteen cells belong to \texttt{dffr} and \texttt{dinf}, which are near-constant at weekly frequency, so a no-change benchmark is near-optimal for them by construction. Among the three series that move continuously, the pipeline beats Zero in 6 of 9 cells: two of three for market returns, one of three for the WEI and three of three for the MES proxy.

\begin{table}[!ht]
\centering
\caption{Joint G3++$\to$VARX OOS RMSE versus Zero}
\label{c2:tab:phase9_rmse}
\small
\resizebox{\textwidth}{!}{%
\begin{tabular}{l cc cc cc}
\toprule
& \multicolumn{2}{c}{$h=4$w} & \multicolumn{2}{c}{$h=12$w} & \multicolumn{2}{c}{$h=26$w} \\
\cmidrule(lr){2-3}\cmidrule(lr){4-5}\cmidrule(lr){6-7}
Variable & G3++$\to$VARX & Zero & G3++$\to$VARX & Zero & G3++$\to$VARX & Zero \\
\midrule
\texttt{mkt\_ret}      & 0.023008 & 0.022477 & 0.026839 & 0.027539 & 0.016794 & 0.017146 \\
\texttt{dffr}          & 0.001543 & 0.001555 & 0.001720 & 0.001739 & 0.000050 & 0.000000 \\
\texttt{dwei}          & 0.002479 & 0.002458 & 0.008066 & 0.008077 & 0.005777 & 0.005772 \\
\texttt{dinf}          & 0.001527 & 0.001526 & 0.000123 & 0.000000 & 0.004146 & 0.004127 \\
\texttt{d\_mes\_proxy} & 0.001227 & 0.001254 & 0.002274 & 0.002294 & 0.000966 & 0.000974 \\
\bottomrule
\end{tabular}}
\begin{flushleft}
\footnotesize The two Zero cells of exactly 0.000000 (\texttt{dffr} at 26w, \texttt{dinf} at 12w) reflect a realised change of zero at every origin; see Section~\ref{c2:subsec:oos_design}. The pipeline has lower RMSE than Zero in 8 of 15 cells.
\end{flushleft}
\end{table}

The question that matters is how much information the curve carries. Table~\ref{c2:tab:skill_decomp_compact} isolates the marginal value of yield-curve conditioning by comparing Mode~B (no curve input), Mode~A (the realised future curve) and the joint pipeline (the simulated curve). If the curve carried exploitable predictive content for these weekly variables, Mode~A should beat Mode~B consistently. It does not: Mode~A is better in only 5 of 15 cells, and in several cells the realised future curve makes the forecast worse. This is not an artefact of the two near-constant series, since among the three continuously moving variables Mode~A wins 3 of 9, the same proportion.

The joint pipeline beats Mode~B in 8 of 15 cells with one tie, which on a count alone might suggest the conditioning helps. The magnitudes say otherwise. Mode~A represents the most the curve could contribute, so the Mode~B to Mode~A gap is the improvement available; that gap exists in only five cells, and in them the pipeline closes more than half of it in two (\texttt{dffr} at 4 weeks and the MES proxy at 12 weeks) and less than a third in the other three. Elsewhere the pipeline's improvements over Mode~B are in the fifth or sixth decimal place. The conditioning moves the forecasts, but not by an amount that matters.

\begin{table}[!ht]
\centering
\caption{Information-Value Decomposition: RMSE Comparison}
\label{c2:tab:skill_decomp_compact}
\small
\resizebox{\textwidth}{!}{%
\begin{tabular}{l ccc ccc ccc}
\toprule
& \multicolumn{3}{c}{$h=4$w} & \multicolumn{3}{c}{$h=12$w} & \multicolumn{3}{c}{$h=26$w} \\
\cmidrule(lr){2-4}\cmidrule(lr){5-7}\cmidrule(lr){8-10}
Variable & B & A & G3++ & B & A & G3++ & B & A & G3++ \\
\midrule
\texttt{mkt\_ret}      & 0.022521 & 0.023452 & 0.023008 & 0.026922 & 0.026666 & 0.026839 & 0.016634 & 0.019630 & 0.016794 \\
\texttt{dffr}          & 0.001558 & 0.001541 & 0.001543 & 0.001738 & 0.001625 & 0.001720 & 0.000043 & 0.000158 & 0.000050 \\
\texttt{dwei}          & 0.002519 & 0.002529 & 0.002479 & 0.008089 & 0.008157 & 0.008066 & 0.005774 & 0.005848 & 0.005777 \\
\texttt{dinf}          & 0.001527 & 0.001544 & 0.001527 & 0.000054 & 0.000294 & 0.000123 & 0.004130 & 0.004161 & 0.004146 \\
\texttt{d\_mes\_proxy} & 0.001233 & 0.001180 & 0.001227 & 0.002280 & 0.002279 & 0.002274 & 0.000994 & 0.001046 & 0.000966 \\
\bottomrule
\end{tabular}}
\begin{flushleft}
\footnotesize B: VARX with exogenous PCs set to zero. A: VARX conditioned on realised future PC paths. G3++: joint G3++$\to$VARX pipeline. Mode~A has lower RMSE than Mode~B in 5 of 15 cells; the joint pipeline has lower RMSE than Mode~B in 8 of 15 cells, with one tie (\texttt{dinf} at 4w).
\end{flushleft}
\end{table}

The conclusion is that, once a variable's own lags are controlled for, the yield curve is not an incremental predictor of weekly first-differenced macro-financial variables. The density evaluation below shows that the same holds for the full predictive distribution.

\FloatBarrier

\phantomsection\label{c2:subsec:results_density}
Point accuracy is only half of what a scenario engine has to deliver; the other half is a correctly calibrated distribution. %

\begin{figure}[!ht]
\centering
\includegraphics[width=0.85\textwidth]{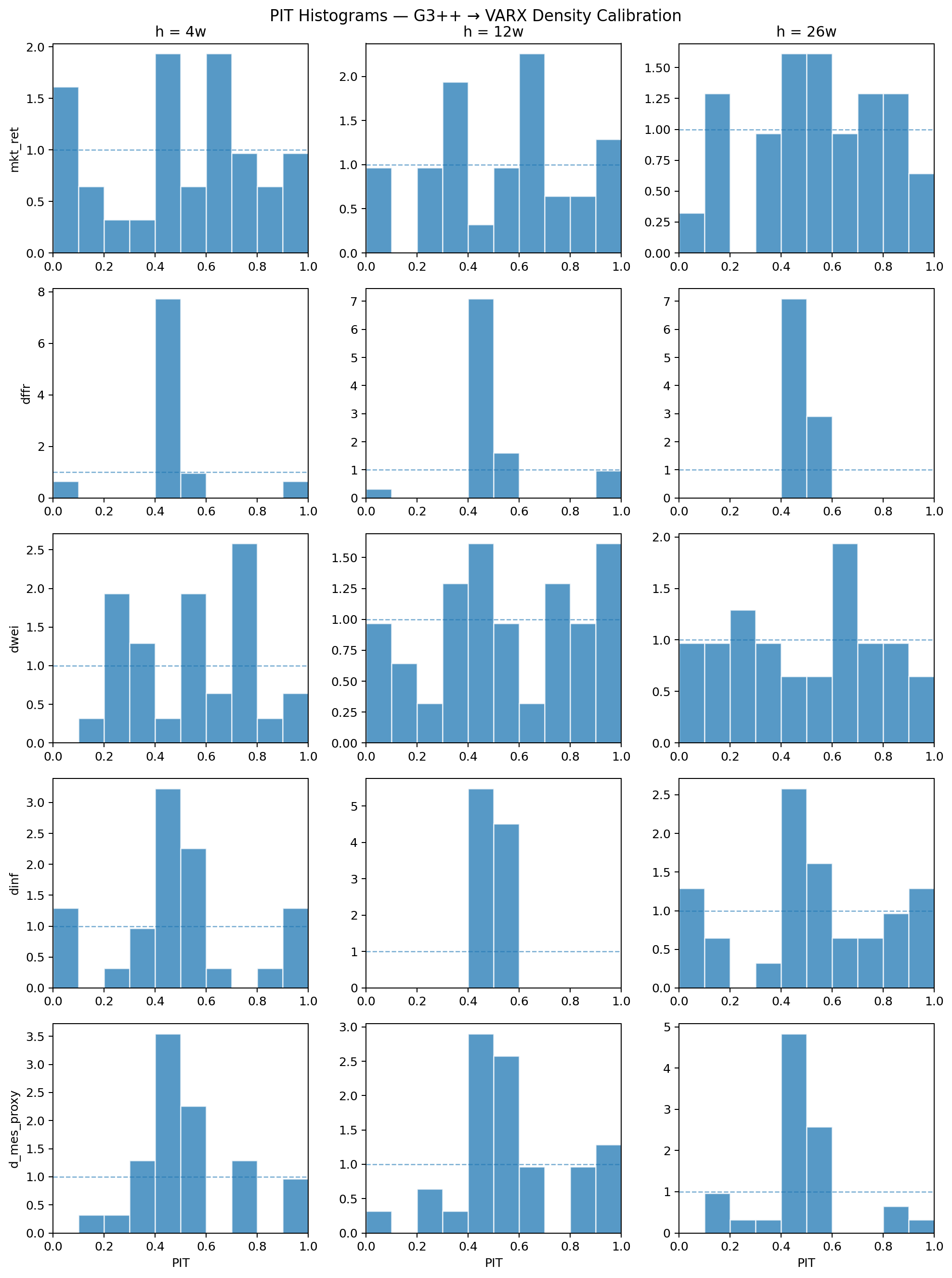}
\caption{PIT histograms for the joint G3++$\to$VARX density forecasts. Dashed line: uniform benchmark.}
\label{c2:fig:pit_histograms}
\end{figure}

Figure~\ref{c2:fig:pit_histograms} and Table~\ref{c2:tab:coverage} assess whether the pipeline's predictive distributions are correctly calibrated. Coverage is close to the nominal 80\% for market returns at 4 and 12 weeks and for the WEI at 12 and 26 weeks. Two cells among the continuously moving series reject correct coverage, both in the conservative direction: the WEI at 4 weeks (93.5\%, $p=0.033$) and the MES proxy at 26 weeks (96.8\%, $p=0.005$). The 100\% coverage for \texttt{dffr} at 26 weeks and \texttt{dinf} at 12 weeks is the mechanical consequence noted in Section~\ref{c2:subsec:oos_design}. For the three continuously moving variables the distributions are therefore well calibrated in seven of nine cells and conservative in the other two; for the two intermittently moving variables the intervals are conservative and sometimes too wide.

\begin{table}[!ht]
\centering
\caption{Kupiec Unconditional Coverage Tests for 80\% Nominal Forecast Intervals}
\label{c2:tab:coverage}
\small
\begin{tabular}{lccccc}
\toprule
Variable & Horizon & Empirical Cov.\ (\%) & $N$ & LR stat & $p$-value \\
\midrule
\texttt{mkt\_ret}      & 4w  & 74.2 & 31 & 0.61 & 0.434 \\
\texttt{mkt\_ret}      & 12w & 80.6 & 31 & 0.01 & 0.928 \\
\texttt{mkt\_ret}      & 26w & 90.3 & 31 & 2.44 & 0.118 \\
\texttt{dffr}          & 4w  & 87.1 & 31 & 1.08 & 0.298 \\
\texttt{dffr}          & 12w & 87.1 & 31 & 1.08 & 0.298 \\
\texttt{dffr}          & 26w & 100.0 & 31 & 13.83 & $<0.001^{***}$ \\
\texttt{dwei}          & 4w  & 93.5 & 31 & 4.55 & $0.033^{**}$ \\
\texttt{dwei}          & 12w & 74.2 & 31 & 0.61 & 0.434 \\
\texttt{dwei}          & 26w & 83.9 & 31 & 0.31 & 0.580 \\
\texttt{dinf}          & 4w  & 74.2 & 31 & 0.61 & 0.434 \\
\texttt{dinf}          & 12w & 100.0 & 31 & 13.83 & $<0.001^{***}$ \\
\texttt{dinf}          & 26w & 74.2 & 31 & 0.61 & 0.434 \\
\texttt{d\_mes\_proxy} & 4w  & 90.3 & 31 & 2.44 & 0.118 \\
\texttt{d\_mes\_proxy} & 12w & 83.9 & 31 & 0.31 & 0.580 \\
\texttt{d\_mes\_proxy} & 26w & 96.8 & 31 & 7.77 & $0.005^{***}$ \\
\bottomrule
\end{tabular}
\begin{flushleft}
\footnotesize $^{**}$ and $^{***}$: rejection of correct unconditional coverage at 5\% and 1\%. All rejections are in the direction of overcoverage. The two 100\% cells reflect a realised change of zero at every origin; see Section~\ref{c2:subsec:oos_design}.
\end{flushleft}
\end{table}

Calibration tests are absolute: they say whether a distribution is right, not whether it is better than an alternative. Table~\ref{c2:tab:crps} therefore reports the CRPS \citep{GneitingRaftery2007} of the joint pipeline against two benchmark distributions over the same horizons, Mode~B and a univariate AR(1) with Gaussian innovations at the training-residual variance, for the three continuously moving variables. Lower is better.

The joint pipeline has the lowest CRPS in two of nine cells, Mode~B in six and the AR(1) in one, but the counts matter less than the magnitudes: the pipeline and Mode~B scores differ by less than 2\% in eight of nine cells, and the largest gap between any two scores in the table is 4.4\% (market returns at 26 weeks). Over 31 origins, differences of that size are unlikely to be statistically distinguishable, so the three predictive distributions are indistinguishable in accuracy.

Mode~B is the informative comparison, because it is the same VARX with only the curve input switched off. That its scores match the pipeline's to within two percentage points means the conditioning contributes essentially nothing to density accuracy, which is the same conclusion the information-value decomposition above reached for point accuracy. Together with the Mode~A comparison, this makes three independent tests of the same proposition, and all three agree.

This result locates the framework's value precisely. It is not a more accurate forecaster, in point or density terms, than the same VARX without curve inputs. What it provides, and what neither Mode~B nor any univariate benchmark can, is a distribution conditional on a \emph{specified} curve path. Mode~B has no way to impose a parallel shift, an inversion or a steepener, because it has no yield-curve input to impose them on. The stress scenarios that close this section illustrate that use.

\begin{table}[!ht]
\centering
\caption{Continuous Ranked Probability Score, Joint Pipeline versus Benchmarks}
\label{c2:tab:crps}
\small
\resizebox{\textwidth}{!}{%
\begin{tabular}{l ccc ccc ccc}
\toprule
& \multicolumn{3}{c}{$h=4$w} & \multicolumn{3}{c}{$h=12$w} & \multicolumn{3}{c}{$h=26$w} \\
\cmidrule(lr){2-4}\cmidrule(lr){5-7}\cmidrule(lr){8-10}
Variable & G3++ & Mode B & AR(1) & G3++ & Mode B & AR(1) & G3++ & Mode B & AR(1) \\
\midrule
\texttt{mkt\_ret}      & 0.012481 & 0.012545 & 0.012540 & 0.013590 & 0.013414 & 0.013515 & 0.009764 & 0.009352 & 0.009626 \\
\texttt{dwei}          & 0.001477 & 0.001492 & 0.001491 & 0.003801 & 0.003743 & 0.003754 & 0.002938 & 0.002906 & 0.002904 \\
\texttt{d\_mes\_proxy} & 0.000639 & 0.000628 & 0.000643 & 0.001011 & 0.001001 & 0.001009 & 0.000567 & 0.000565 & 0.000569 \\
\bottomrule
\end{tabular}}
\begin{flushleft}
\footnotesize Mean CRPS across 31 origins; lower is better. Mode~B is the same VARX with exogenous PC changes set to zero. AR(1) uses Gaussian innovations at the training-residual variance. Lowest score per cell: G3++ in 2 of 9, Mode~B in 6, AR(1) in 1. G3++ and Mode~B differ by less than 2\% in 8 of 9 cells; the largest gap between any two scores is 4.4\% (\texttt{mkt\_ret}, 26w).
\end{flushleft}
\end{table}

\FloatBarrier

\phantomsection\label{c2:subsec:results_stress}

\begin{figure}[!ht]
\centering
\includegraphics[width=\textwidth]{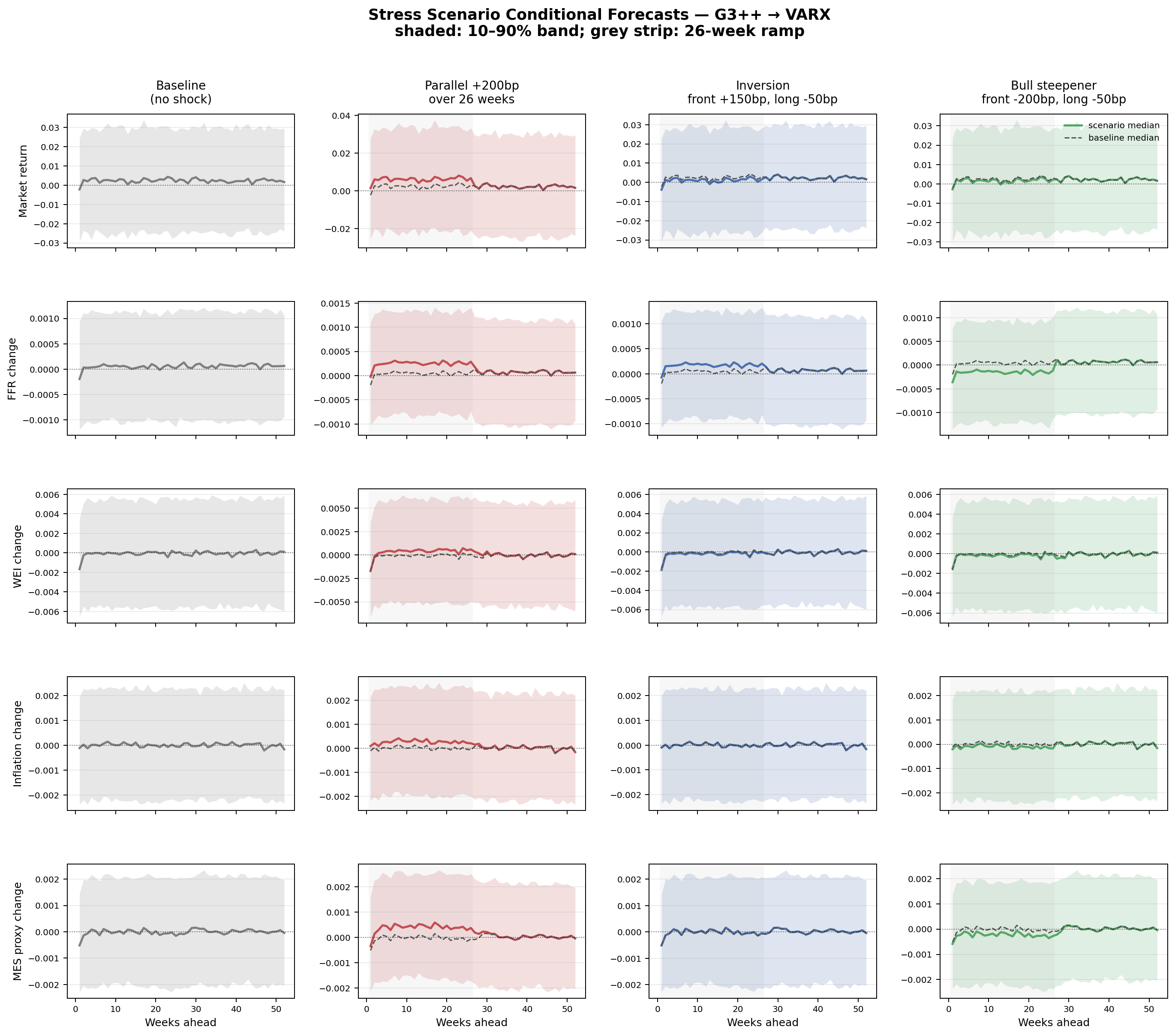}
\caption{Stress-scenario conditional forecasts. Columns: no-shock baseline, parallel $+200$\,bp, inversion, bull steepener. Shading: 10--90\% bands; grey strip: 26-week ramp; dashed line: baseline median.}
\label{c2:fig:stress_scenarios}
\end{figure}

Three curve paths are imposed from the December 2024 anchor, each reached over 26 weeks and then held: a \emph{parallel $+200$\,bp} rise; a \emph{curve inversion} (front end $+150$\,bp, long end $-50$\,bp); and a \emph{bull steepener} (front end $-200$\,bp, long end $-50$\,bp). Figure~\ref{c2:fig:stress_scenarios} reports the 10--90\% fan charts and medians over 52 weeks, and Table~\ref{c2:tab:stress} reports the median responses during the ramp as differences from the no-shock baseline, together with their cumulative size over the 26 weeks.

The three paths produce distinct distributions, and the differences run in the directions the sample's history implies. During the parallel rise the median weekly equity return is 0.37 percentage points above baseline, which cumulates to about 9.7 percentage points over the ramp; the funds rate rises by a further 54\,bp in total, the WEI by 1.2 points, year-on-year inflation by 0.7 percentage points and the MES proxy by 1.1 percentage points. The inversion, in which the front end rises while the long end falls, is the one that resembles a tightening: the funds rate rises by 34\,bp in total and equity returns run 0.13 percentage points a week below baseline, cumulating to about $-3.5$ percentage points. The bull steepener runs the other way, with the funds rate 49\,bp lower, inflation 0.3 percentage points lower and the MES proxy 0.5 percentage points lower over the ramp, and a small equity loss. Once the ramp ends and the curve stops moving, every median returns to the baseline within a week or two, because the VARX responds to factor changes and its own dynamics have a half-life of little more than a week. All three shifts are modest relative to dispersion: the 80\% bands for weekly equity returns are about six percentage points wide in every scenario, so the scenarios move the centre of the distribution rather than its width.

Two points follow for how the exercise is used. First, the scenarios return the historical association between curve movements and macro-financial outcomes as learned by the VARX, not the causal effect of a policy decision. In the sample, parallel rises in the curve accompanied growth and reflation, which is why the parallel scenario raises equity returns, activity and inflation together; a supervisor who wants the effect of a tightening surprise should combine the curve path with the sign-identified shock reported at the start of this section, which restricts the equity response directly. Second, the exercise does what no univariate or no-change benchmark can. A curve path stated in policy terms is converted into a joint distribution for activity, inflation, equity returns and bank tail risk with the same dispersion the model produces unconditionally, which is the object a stress test needs and the reason the conditioning channel exists.

\begin{table}[!ht]
\centering
\caption{Stress Scenarios: Median Responses During the 26-Week Ramp, Relative to the No-Shock Baseline}
\label{c2:tab:stress}
\small
\setlength{\tabcolsep}{4pt}
\begin{tabular}{l rr rr rr}
\toprule
& \multicolumn{2}{c}{Parallel $+200$\,bp} & \multicolumn{2}{c}{Inversion} & \multicolumn{2}{c}{Bull steepener} \\
\cmidrule(lr){2-3}\cmidrule(lr){4-5}\cmidrule(lr){6-7}
Variable & weekly & cum. & weekly & cum. & weekly & cum. \\
\midrule
Equity return (pp)          & $+0.37$ & $+9.7$  & $-0.13$ & $-3.5$ & $-0.07$ & $-1.9$ \\
Funds-rate change (bp)      & $+2.1$  & $+54$   & $+1.3$  & $+34$  & $-1.9$  & $-49$  \\
WEI change (points)         & $+0.05$ & $+1.2$  & $-0.01$ & $-0.3$ & $-0.01$ & $-0.3$ \\
Inflation change (pp, YoY)  & $+0.03$ & $+0.7$  & $0.00$  & $0.0$  & $-0.01$ & $-0.3$ \\
MES proxy change (pp)       & $+0.04$ & $+1.1$  & $0.00$  & $0.0$  & $-0.02$ & $-0.5$ \\
\bottomrule
\end{tabular}
\begin{flushleft}
\footnotesize Median path minus the no-shock baseline median, averaged over weeks 1 to 26 (weekly) and summed over them (cum.). Baseline and scenarios share the same random numbers. After week 26 the differences are zero to the precision shown.
\end{flushleft}
\end{table}

\FloatBarrier

\section{Policy Recommendations}
\label{c2:sec:policy}

Five recommendations follow from the results.

\paragraph{Report the measure and the implied risk premium.} A term-structure model used for scenario generation does two jobs. Pricing must be risk-neutral for the curves to be arbitrage-free; simulation must be physical for them to describe how yields move; and the market price of risk that links the two is where the model's term premium lives. Table~\ref{c2:tab:lambda_implied} reports it in three numbers. An institution that cannot produce the equivalent table for its own generator does not know what term premium its scenarios embed, and a supervisor reviewing such a model should ask for it.

\paragraph{Test the yield curve's predictive content at the frequency you use.} The link between the term structure and future activity is robust at quarterly and annual horizons. It does not follow that it holds weekly, and here it does not: conditioning on the realised future curve beats no conditioning in only 5 of 15 cells, and the predictive densities are indistinguishable. An institution building a high-frequency monitoring dashboard around curve movements should run the Mode~A comparison of Section~\ref{c2:subsec:results_phase9a} at its own frequency first. It is cheap and needs no new data.

\paragraph{Treat a tail-risk level measure and a co-movement measure as complements.} MES and the bank--market covariance proxy respond in opposite directions to the same tightening shock, with the entire 80\% band on each side of zero. MES falls because tightening lowers all equity prices; the covariance proxy rises because bank-sector returns move more closely with the market. A framework reporting only one of them will register tightening as reducing systemic risk or as increasing it, depending purely on which was chosen. Reporting both, and reading the divergence, is the more informative practice.

\paragraph{Specify scenarios as paths, and distinguish a path from a policy shock.} A stress scenario stated as a level of the curve conveys nothing to a transmission layer that responds to movements; it has to be stated as a path with a pace, and the pace is part of the scenario. Imposing a path then returns the historical comovement associated with such moves, not the causal effect of a policy decision. The parallel $+200$\,bp ramp raises equity returns, activity and inflation together, because in the sample that is what accompanied parallel rises; read as a tightening scenario it would be misleading, while the inversion is the path that behaves like one. Where the question is about policy, the curve path should be combined with the sign-identified shock, and stress-testing practice should label the two outputs separately.

\paragraph{Evaluate scenario engines against simple benchmarks.} RAMSI and BEAST are not publicly evaluated against random-walk, AR(1) or no-change baselines. The apparatus required is modest: an expanding window, Diebold--Mariano tests for point accuracy, and PIT, coverage and CRPS diagnostics for densities. Applied here, it produced a result the in-sample evidence would not have revealed. A supervisor commissioning a scenario engine should ask for this evaluation as a matter of course.

Two cautions apply throughout. The findings rest on one system, one currency, weekly first differences and a post-2010 sample, so the result on incremental predictive content should be re-established rather than assumed at other frequencies or in other markets. And the impact magnitudes in Section~\ref{c2:subsec:results_irf} are set-identified, normalised and orthogonal to the contemporaneous curve movement, so the sign pattern and the cross-indicator comparison carry the interpretation, not the scale.

\section{Conclusion}
\label{c2:sec:conclusion}

This paper set out to build a scenario engine that converts a yield-curve path into a conditional macro-financial distribution, and to evaluate it to the standard the forecasting literature applies. Both are contributions, and the second is what gives the first its weight.

The term-structure block does what a scenario generator needs to do. It reproduces the observed discount curve exactly, matches weekly zero-rate volatility across the 1Y--10Y segment, and has the lowest out-of-sample RMSE in 13 of 20 maturity--horizon cells with a tie in a fourteenth, the clearest gains being at the 1Y, 2Y and 5Y pillars where policy expectations dominate. The methodological step that makes this possible is keeping the two probability measures distinct: pricing under $\mathbb{Q}$ preserves arbitrage consistency, simulating under $\mathbb{P}$ reproduces how yields move, and the market price of risk that links them is estimated and reported. Table~\ref{c2:tab:lambda_implied} puts the term premium embedded in every scenario on the page as three numbers, none exceeding 0.09 in absolute value. Applied work in this area frequently leaves the choice of measure unstated; stating it costs nothing and makes the economic content of the generator visible.

The structural block delivers an interpretable transmission result. A contractionary policy shock is associated with falling equity returns, activity, inflation and the MES proxy on impact, and the relevance filter and magnitude caps are what make those responses usable at weekly frequency. The most informative single result is the comparison between MES and the CoVaR-inspired covariance proxy: the two move in opposite directions under the same shock, MES pushed down through the discount-rate channel while the covariance proxy rises, consistent with the risk-taking channel of \citet{BorioZhu2012}. That divergence matters for anyone using either measure as a systemic-risk indicator, because it shows the two are not substitutes.

The evaluation produces a finding of independent interest. Three separate tests, the Mode~A comparison, the joint-pipeline decomposition and the CRPS comparison, agree that at weekly frequency, once each variable's own lags are controlled for, the term structure carries no measurable incremental information about these macro-financial variables, in the conditional mean or in the full predictive distribution. Conditioning on the realised future curve beats no conditioning in only 5 of 15 cells; the pipeline, the same VARX without curve inputs and a univariate AR(1) are within two per cent of one another in CRPS in eight of nine cells. The result holds for this system, on first-differenced weekly variables over a post-2010 sample, and is not a claim about the yield curve at lower frequencies or in levels. It is a useful result for anyone building high-frequency monitoring around the curve, and it locates the framework's value precisely.

That value lies in the conditioning channel. Section~\ref{c2:subsec:results_stress} shows three curve paths producing three distinct joint distributions, each read against a common baseline, with the parallel rise, the inversion and the steepener moving equity returns, the funds rate, activity, inflation and bank tail risk in the directions the sample's history implies. A random walk, an AR(1) or a no-change forecast cannot take a stipulated curve path and return anything, because none has a yield-curve input on which such a path could be imposed. This architecture can, and does so while remaining arbitrage-free, transparent and reproducible. That is the capability stress testing requires, where the question is the range of plausible outcomes under a specified interest-rate environment rather than the most accurate unconditional forecast, and it is a capability the institutional frameworks closest to this one, RAMSI and BEAST, do not expose to public evaluation.

Several extensions follow. The price of risk is held constant, which places the term-structure block in the completely affine class; an essentially affine specification in the sense of \citet{Duffee2002} is the natural next step. The Gaussian three-factor structure underfits the ultra-long end and abstracts from regime shifts, stochastic volatility and jump risk. Two of the five endogenous variables move only intermittently at weekly frequency, which argues for a nowcast-based inflation series or a mixed-frequency specification. The identified policy shock is orthogonal to the contemporaneous curve movement by construction, and a specification in which the curve responds to the shock would recover the part of transmission that the exogenous block absorbs. The MES proxy captures short-run market-implied tail-risk pricing rather than the slower balance-sheet channel, so funding-spread or CDS-based measures would complement it, and replacing the covariance proxy with a full quantile-regression CoVaR would sharpen the co-movement comparison. Testing the framework in non-U.S. monetary regimes, or embedding it upstream of a bank-level supervisory model, would extend it further.

What the paper establishes is that a pricing model can be repurposed into a physical-measure scenario generator with a stated risk premium, coupled to a transparent transmission layer, and evaluated to the standard the forecasting literature demands. The architecture is the contribution, and the evaluation is what makes it usable.

\section*{Declarations}

\paragraph{Funding.} Ahmad Qadi was supported by a doctoral scholarship from the Saudi Arabian Ministry of Education. This research did not receive any other specific grant from funding agencies in the public, commercial or not-for-profit sectors.

\paragraph{Competing interests.} The authors declare that they have no known competing financial interests or personal relationships that could have appeared to influence the work reported in this paper.

\paragraph{Data availability.} The zero-coupon yield curves are licensed from the London Stock Exchange Group and cannot be redistributed. The macro-financial series are public and available from the Federal Reserve Economic Data (FRED) database and the Federal Reserve Bank of Dallas. The analysis code is available from the corresponding author on request; licensed data are excluded.

\paragraph{Use of generative AI.} During the preparation of this work, the authors used Claude (Anthropic) and ChatGPT (OpenAI) to assist with parts of the text, check reported numbers against tables and code output, verify bibliographic details and debug LaTeX and code. All formulas, empirical claims, references, notation and manuscript content were reviewed and verified by the authors, who take full responsibility for the content of this manuscript.

\end{document}